\documentstyle[12pt,a4,amssymb,times]{article}

\def\a{\alpha}
\def\pa{\alpha^\prime}
\def\b{\beta}
\def\pb{\beta^\prime}
\def\th{\theta}
\def\S{\Sigma}

\def\f{\phi}
\def\F{\Phi}
\def\vf{\varphi}
\def\D{\Delta}
\def\e{\epsilon}

\def\c{\chi}
\def\G{\Gamma}
\def\g{\gamma}
\def\o{\omega}

\def\p{\pi}
\def\d{\delta}

\def\l{\lambda}
\def\L{\Lambda}
\def\pL{\Lambda^\prime}

\def\pri{i^\prime}
\def\pj{j^\prime}

\def\i{\iota}

\def\ct{{\cal T}}
\def\cd{{\cal D}}
\def\cj{{\cal J}}
\def\cv{{\cal V}}
\def\ca{{\cal A}}
\def\cg{{\cal G}}

\def\half{\frac12}
\def\pr{\partial}
\def\to{\rightarrow}

\newcommand{\be}{\begin{equation}}
\newcommand{\ee}{\end{equation}} 
\newcommand{\bea}{\begin{eqnarray}}
\newcommand{\eea}{\end{eqnarray}}

\def\bra#1{\left\langle#1\right|}
\def\ket#1{\left|#1\right\rangle}

\begin{document}

\begin{titlepage}

\begin{flushright} 
IST/DM
\end{flushright}

\bigskip
\bigskip
\bigskip
\bigskip

\begin{center}

{\bf{\Large Quantum Field Theory of Spin Networks }}

\end{center}
\bigskip
\begin{center}
 A. Mikovi\'c \footnote{E-mail address: amikovic@math.ist.utl.pt. On leave of 
absence 
from Institute of Physics, P.O.Box 57, 11001 Belgrade, Yugoslavia}
\end{center}
\begin{center}
\it
Departamento de Matem\'atica and Centro de Matem\'atica Aplicada, 
Instituto Superior 
Tecnico, Av. Rovisco Pais,
1049-001 Lisboa, Portugal

\end{center}

\normalsize 

\bigskip 
\bigskip
\begin{center}
                        {\bf Abstract}
\end{center}    
We study the transition amplitudes in the state-sum models of quantum gravity
in $D=2,3,4$ spacetime dimensions by using the field theory over $G^D$ 
formulation, where $G$ is the relevant Lie group. By promoting the group 
theory Fourier modes into creation and 
annihilation operators we construct a Fock space for the quantum field theory
whose Feynman diagrams give the transition amplitudes.
By making products of the Fourier modes we construct operators and states 
representing the spin networks 
associated to triangulations of spatial boundaries of a triangulated
spacetime manifold. The corresponding spin network amplitudes give
the state-sum amplitudes for triangulated manifolds with boundaries.
We also show that
one can introduce a discrete time evolution operator, where
the time is given by the number of $D$-simplices in a 
triangulation, or equivalently by the number of the vertices of the Feynman 
diagram. The corresponding transition amplitude is a finite sum of Feynman 
diagrams, and in this way one avoids the problem of infinite amplitudes 
caused by summing over all possible triangulations. 
\end{titlepage}
\newpage

\section{Introduction}

The idea of constructing a quantum theory of gravity by assuming that
the spacetime is discrete at the Planck scale is appealing
because it resolves by definition the problem of UV infinities of the 
conventional quantum field theory approach. However, the development of
this idea was hindered by the complexity of the obvious candidate
for such a theory, i.e. Regge simplicial formulation of general relativity 
(GR) \cite{re}. 

In the past decade a new class of simplical gravity models have been 
developed, which are based on the representation theory of the relevant 
symmetry group $G$. This group is 
$SO(D)$ for the Euclidean gravity case or $SO(D-1,1)$ for the Lorentzian case.
The prototype model was the Ponzano-Regge model of 3d gravity \cite{por}, which
was made mathematically well defined in the work of Turraev and Viro by
passing from the $SO(3)$ to the quantum $SU(2)$ group \cite{tv}. This was 
generalized to 4d in the approach of  topological state 
sum models \cite{cy}, which is a mathematical formalism for constructing 
topological invariants for manifolds out of colored triangulations, based
on the category theory. In physics terms, this is a way of 
constructing partition functions for topological field theories.
The state-sum formalism was based on the papers of Boulatov and Ooguri, who 
realized that the topologically invariant state sums can be generated as 
Feynman
diagrams of a field theory over $G^D$ \cite{b,o}. Their approach was 
motivated by the
result from string theory that the partition function of a discretized
string theory can be represented as a sum of Feynman diagrams of a matrix 
field theory (for a review and references see \cite{stmm}). 

Based on these developments, Barret and Crane have proposed an approach for 
obtaining the partition 
function for general relativity (GR) in $D=4$, which is a
non-topological theory, from a state sum of a topological theory, 
the BF theory \cite{bce,bcl}. The BC model represents a very 
interesting formulation of discrete quantum gravity, since the local
symmetry group of the spacetime plays a crucial role. One colors a
triangulation $\ct$ of the spacetime manifold $M$ with a certain class of 
irreducible 
representations (irreps) of the Lorentz group, and the partition function
for a given $\ct$ is the sum over the amplitudes for all colorings. 
Given the partition function, one can compute the transition
amplitude associated with a triangulation of $M$ with spatial boundaries 
which are
colored with a fixed set of irreps, by summing over the internal irreps,
exactly as in the topological cases \cite{por,fr,baez}.

Since a field theory formulation of the BC model exists \cite{dfkr}, such an 
amplitude would correspond to a Feynman diagram with external legs.
In the quantum field theory (QFT) formalism such
diagrams  contribute to the 
matrix elements of an evolution operator, the S-matrix. This evolution
operator acts in the Hilbert space of the QFT, which is the Fock space 
constructed out of the creation and the annihilation operators. The aim of this
paper is to develop these concepts for the BC model. 

A further motivation is to 
construct the states which correspond to the colored triangulations of 
the spatial boundaries, i.e. the spin network states. This is based on the
property of the QFT formalism that one can associate states to operators via
the Fock space vacuum state, so that one could in principle construct
the operators corresponding to spin networks. This would be also the first 
step for the third quantization formulation, i.e. a quantum theory where the 
number of the universes is not fixed. 

Furthermore, one can define the transition amplitudes
between these spin net states by summing over
the appropriate Feynman diagrams with external legs. Since such a 
Feynman diagram (FD)
corresponds to a triangulation of the spacetime with boundaries, there are 
several amplitudes one can 
associate to a transition from a state on the initial surface to a state on 
the final surface. The standard amplitude is given by the sum over all 
possible triangulations, or equivalently over all FD, but we argue here that 
it makes sense to consider
the perturbative part of that amplitude, which is given by the sum of the
FD with $n$ vertices, or triangulations with $n$ simplices, as the transition 
amplitude for the time interval of $n$ Planck units of time. In this way one 
introduces a discrete time variable, and also regularizes the amplitude, since
one avoids the infinite sum over all triangulations, which is a generically 
divergent expression.     

In sections 2,3 and 4 we study the simplical field theory for a 2d 
discretized
gravity theory based on the 2d BF theory, since it is a good toy model for 
illustrating and developing our ideas. In section 5 we further develop and
explore the general concepts introduced in the case of the 2d model, where we
consider the 3d simplical field theory of Boulatov. In 
sections 6 and 7 we apply these concepts to the field theory formulation of
the BC model. In section 8 we present our conclusions.  

\section{D=2 model}

We start with the $D=2$ model, since it is very useful for understanding and
illustrating the basic concepts. The $D$-dimensional BF theory is given by 
the action
\be
S_{BF}= \int_M <B \wedge F > \label{bf}
\ee
where B is a $(D-2)$-form,i.e. scalar field in 2d, and $F = dA + A\wedge A$ is 
the curvature two-form for the connection one-form $A$. $A$ and $B$ 
take values in the Lie algebra of the Lie group $G$ and $<,>$ is the 
corresponding invariant bilinear form.

For the case of 2d gravity the relevant groups are the 2d Poincare group 
$ISO(1,1)$, then the 2d anti-de-Sitter
group $SO(1,2)$ and its Euclidean versions $SO(3)$ and $SU(2)$. 
The corresponding BF theory 
gives the Jackiw-Taitelboim theory \cite{jt}, which follows from the 
identifications
\be
A = \o J_0 + e^\pm J_\pm 
\quad,\quad B= B^0 J_0 + B^\pm J_\pm 
\ee
where $\o$ is the spin connection, $e^\pm$ are the zweibeins, $B^0$
is the dilaton, and $J$ are the Lie algebra generators, so that
\be 
S_{BF} = \int_M B^0 (d\o + \l_c e^+ \wedge e^- )
+ B^\mp (de^\pm \pm \o e^\pm ) \,,  
\ee 
where $\l_c$ is the cosmological constant. Hence the $B^\pm$ enforce the
zero-torsion conditions which give $\o = \o (e)$ and one ends up with the 
Jackiw-Teitelboim action. We will consider the compact $G$ case, i.e. 
Euclidean 2d metrics, since the formulas are simpler, and the basic ideas are 
the same.

The field theory which generates the partition functions for the 
triangularizations of the two-manifold $M$ as Feynman diagrams is
given by
\be
S_2 = \half\int_{G^2}d^2 g\, \vf^2 (g_1 ,g_2) 
+ {\l\over 3!}\int_{G^3}
d^3 g \,\vf (g_1 ,g_2)\vf (g_2 ,g_3)\vf (g_3 ,g_1)\quad,\label{tda}
\ee
where $\vf$ satisfies
\be
\vf (g_1 , g_2 ) = \vf (g_1 g , g_2 g) \quad,
\ee
and $\l$ is the perturbation theory expansion parameter.
Hence
\be
\vf (g_1 , g_2 ) = \vf (g_1 \cdot g_2^{-1} ) = \sum_{\L,\a,\b} \f^\L_{\a\b} 
D^\L_{\a\b} (g_1 \cdot g_2^{-1}) \quad,\label{fex}
\ee
where the last formula follows from the Peter-Weyl theorem.
$\L$ denotes an unitary irreducible representation of $G$, $1\le \a,\b \le
d_\L = $ dim$ \L$. The Fourier coefficients $\f^\L$ will be important for the
construction of the spin-network states and transition amplitudes, and they 
are the analogs of the creation and the annihilation operators from the 
particle field theory.   

By inserting (\ref{fex}) into  (\ref{tda}) and by using the orthonormality
relations
\be
\int_G \,dg  (D^\L_{\a\b} (g))^* D^{\pL}_{\pa,\pb} (g) = 
{1\over d_\L}\d^{\L,\pL} \d_{\a,\pa} \d_{\b,\pb} \,,
\ee
and the complex conjugation relations\footnote{This form is for the $SU(2)$ case, so that $\L = 2j= 0,1,2,...$ and $d_\L = \L + 1$.}
\be
(D^\L_{\a\b} (g))^* = (-1)^{\L -\a -\b} D^\L_{-\a -\b} (g)\,, \label{ccr} 
\ee
we obtain for the kinetic part of $S$
\be
S_k = \half\sum_{\L,\a,\b}d_\L^{-1}(-1)^{\L -\a -\b}\f^\L_{\a\b}\f^\L_{-\a-\b} 
= \half \sum_{\L,\a,\b} d_\L^{-1} |\f^\L_{\a\b}|^2 
\quad,\label{tdk}
\ee
while for the interaction part we get
\be
S_v = {\l\over 3!}\sum_{\L,\a,\b,\g}d_\L^{-2}\f^\L_{\a\b}\f^\L_{\b\g}
\f^\L_{\g\a} \quad.\label{tdi}
\ee 
Note that (\ref{ccr}) induces the following reality condition for the
Fourier components $\f$
\be
(\f^\L_{\a\b})^* = (-1)^{\L -\a -\b} \f^\L_{-\a -\b}
\quad,  \label{fcc} 
\ee
which is a non-Abelian generalization of the reality condition for the
usual Fourier modes: $a_p^* = a_{-p}$. 

As far as the Feynmanology is concerned, it will be useful to consider a
general $\f^3$ theory given by the action
\be
S = \half\sum_{I,J} \f_I \f_J C_{IJ}
+ {\l\over 3!}\sum_{I,J,K} \f_I \f_J \f_K V_{IJK}\quad,\label{fcu}
\ee
where the indices $I,J,K$ belong to a general label set.

The perturbative expansion an the Feynman diagrams are generated 
by using the generating functional
\be
Z [\cj] = \int \cd\f \exp ( i (S[\f] + \sum_I\cj_I \f_I )) \quad,
\ee
where $\cd\f =\prod_I d\f_I$. One can write
\be
Z(\cj) = \sum_I \cj_{I_1}\cdots \cj_{I_n} G_{I_1 \cdots I_n}
\ee
where the Green functions $G$ can be written as
\bea
G_{I_1 \cdots I_n} &=&{i^n \over n!} \int \cd\f e^{i (S_k + S_v)} \f_{I_1}
\cdots\f_{I_n} \nonumber \\
&=&{i^n \over n!} \sum_m {i^{m} \over m!}\int\cd\f e^{i S_k}(S_v)^m \f_{I_1}
\cdots\f_{I_n} \nonumber \\ 
&=&{i^n \over n!} \sum_m {i^{m} \over m!}\langle (S_v)^m \f_{I_1}\cdots
\f_{I_n} \rangle = 
{i^n \over n!}\sum_m G^{(m)}_{I_1\cdots I_n } \quad.   
\eea

The perturbative components $G^{(m)}$ can be calculated by a repeated 
differentiation of the Gaussian integral
\be
Z_0 [\cj] = \int \cd\f \exp \left( i S_k [\f] + i \sum_I \cj_I \f_I \right) = 
\D \exp\left( -i\half\sum_{IJ}\cj_I C^{-1}_{IJ} \cj_J 
\right) 
\ee
with respect to the currents $\cj$, where $\D$ is a constant proportional to 
${\det}^{-\half}||C_{IJ}||$. One can then show that the result for
$G^{(m)}$  can be expressed by using the classical analog of 
Wick's theorem from quantum field theory, i.e.
one makes all possible pairings in the string of $\f$'s  
given by $ S_v\cdots S_v  \f_{I_1} \cdots \f_{I_n}$ and then assigns to
each pairing $(\f_I , \f_J )$ a number equal to $C^{-1}_{IJ}$, and then takes
the product of these numbers and sums over all possible sets of pairings.

Wick's theorem is the basis for the Feynman diagrammatic expansion since
topologically non-equivalent sets of pairings can be denoted as trivalent
graphs, which are the Feynman diagrams. For example, the term in 
$G^{(2)}_{IJKL}$ given by
\be
\G_{IJKL} = \# \sum_{I^\prime,J^\prime,K^\prime,L^\prime,M,N} 
C^{-1}_{II^\prime} C^{-1}_{JJ^\prime} 
C^{-1}_{KK^\prime} C^{-1}_{LL^\prime} C^{-1}_{MN} 
V_{I^\prime J^\prime M} V_{K^\prime L^\prime N}
\ee
where $\#$ is the number of equivalent pairings, will 
correspond to the H diagram. In order to obtain its value, which is the 
corresponding S-matrix
contribution, or the transition amplitude, one has to remove the 
external legs propagator factors. This corresponds
to calculating
\bea
A_{IJKL} &=& \sum_{I^\prime J^\prime K^\prime L^\prime}  
C_{II^\prime} C_{JJ^\prime} C_{KK^\prime} C_{LL^\prime} 
 \G_{I^\prime J^\prime K^\prime L^\prime} \nonumber\\ 
&=& \# \sum_{MN} C^{-1}_{MN} V_{IJM} V_{KLN}  \quad.
\eea 

The partition function is given by  $\langle e^{iS_v}\rangle$ and it is equal
to the sum of all vacuum diagrams. The connected diagrams are generated by 
$W[\cj]=\log Z[\cj]$. If $W = \sum_n W_n \l^n $, then the connected
vacuum diagrams are given by 
\bea
W_2 &=& (theta) + (dumbbell)\nonumber \\
W_4 &=& (Mercedes-Benz) + (cylinder) + \cdots\nonumber \\
W_6 &=& (prism) + \cdots \quad, 
\eea
where 
\bea
(theta) &=& \# V_{IJK}V_{LMN} C^{IL} C^{JM} C^{KN} \nonumber  \\
 (dumbbell)&=& \# V_{IJK}V_{LMN} C^{IJ}C^{KL}C^{MN} \nonumber \\ 
(Merc.-Benz) &=& \# V_{IJK}V_{LMN}V_{OPQ}V_{RST} C^{ON} C^{PR} C^{QK} 
C^{LI} C^{JS} C^{TL} \,, 
\label{mbd}
\eea
and so on, where $C^{IJ}= C^{-1}_{IJ}$ and $\#$ is the corresponding 
combinatorial factor.

Hence the vacuum Feynman diagrams denote the rules of forming invariants by 
contracting the indices in the product of tensor quantities $V$, i.e.
given a product $V_{I_1 J_1 K_1} \cdots V_{I_n J_n K_n}$, the corresponding 
Feynman diagrams represent
all topologically non-equivalent ways of contracting the indices in this 
product 
with $C^{IJ} = C^{-1}_{IJ}$. The matrix $C^{IJ}$ is the propagator in the 
field theory language. If a set of indices $\{I,J,...\}$ remains uncontracted, 
then this gives a Feynman diagram with external legs, 
which contributes to the transition amplitude for a set of 
states carrying the quantum numbers $\{I,J,...\}$.

Consider the $(n+m)$-point Green function
\be
G_{n+m} ={i^{n+m}\over (n+m)!} \langle e^{iS_v}\f_{I_1} \cdots \f_{I_n} \f_{J_1} \cdots \f_{J_m} 
\rangle \,.\label{nmgf}
\ee
It will generate all FD with $(n+m)$ external legs. Note that the subclass
of these FD containing all connected FD which have a string\footnote{By a string we mean that there is no vertex with an internal line between the legs of a string.} of $n$ legs and a string of $m$ legs will contribute to the 
amplitude for a transition from a
state labeled with quantum numbers $I_1,...,I_n$ to the state labeled
with quantum numbers $J_1,...,J_m$. This can be explicitly realized by using 
the concept of creation and annihilation operators from the operator 
quantization formalism. The index label set $I$ can be always split as
$I=\{i,-i,j,-j,k,l \}$ such that
\be
\f^*_{i}= \f_{-i} \quad,\quad \f^*_{j}= -\f_{-j} \quad,\quad 
\f^*_{k}= \f_{k} \quad,\quad \f^*_{l}= -\f_{l} 
\quad.
\ee
Then promote $\f_{\pm i}$ and $\f_{\pm j}$ modes into operators 
${\f}_{\pm i}$ and $\sqrt{-1}{\f}_{\pm j}$ satisfying
\be
\f^{\dagger}_{\tilde i}= \f_{-\tilde i} \quad,\quad 
[\f_{\tilde i}, \f^{\dagger}_{\tilde j}]=\d_{\tilde i,\tilde j} \,,
\ee
where $\tilde i ,\tilde j \in \{i,j\}$. The zero-modes $\f_{k}$ and $\f_{l}$
promote into operators $\frac12(\f_{k} + \f^\dagger_k )$ and 
$\frac12 (\f_{l} - \f^\dagger_l )$ where
\be
[\f_{\tilde k}, \f^{\dagger}_{\tilde l}]=\d_{\tilde k,\tilde l} \quad,\quad
 \quad,\quad
\tilde k , \tilde l \in \{ k,l \}\,,
\ee
and all other commutators are zero.

These operators naturally act on the Hilbert space of states
\be
\f^\dagger_{\pri_1}\cdots \f^\dagger_{\pri_n}\ket{0} \quad,\quad 
\f_{\pri} \ket{0} = 0\,,
\ee
where now $\pri \in \{i,j,k,l \}$. 
Then the amplitude for a transition from the initial state 
\be
\ket{\Psi_1} = \f^\dagger_{\pri_1} \cdots \f^\dagger_{\pri_n} \ket{0}
\ee 
to the final state 
\be
\ket{\Psi_2} = \f^\dagger_{\pj_1} \cdots \f^\dagger_{\pj_m} \ket{0}\quad,
\ee
is given by
\be
A_{12} = \bra{0}\f_{\pj_1} \cdots \f_{\pj_m} \exp(i:S_v:) \f^\dagger_{\pri_1} 
\cdots \f^\dagger_{\pri_n} \ket{0} \quad,
\label{tra}
\ee
where $:S_v:$ is the normal-ordered operator $S_v$ with respect to the vacuum 
$\ket{0}$, and all the subsequent
manipulations are the same as in the particle field theory case, i.e. one
uses the operator formulation of Wick's theorem.

Note that one does not need to go to the operator formalism in order to
compute the amplitudes. One can simply say that there are abstract states 
$\ket{I_1 ... I_n}$ and the transition amplitude from the state
$\ket{J_1 \cdots K_1}$ to the state $\ket{L_2 \cdots M_2}$
will be given by  the sum of the corresponding Feynman diagrams. 
We will denote the result as
\be
A_{12} = \bra{L_2 \cdots M_2} \exp(i S_v )\ket{J_1 \cdots K_1}  \quad.
\label{tra}
\ee
We will also use the notation
\be
A_{12} = \langle \f_{L_2} \cdots \f_{M_2} \exp(i S_v )\f_{J_1} \cdots 
\f_{K_1} \rangle  
\quad,
\label{vtr}
\ee
because it indicates that the transition amplitude is obtained from the Green's
function by taking only the contractions corresponding to 
connected Feynman diagrams with a 
string of $n$ and a string of $m$ external legs, which we will denote as
$(n,m)$ FD.

The 
expression (\ref{vtr}) is calculated perturbatively in $\l$, where at order 
$n$ one has to calculate
\be
A^{(n)}_{12} = \langle \f_{L_2} \cdots \f_{M_2} (S_v)^n \f_{J_1} \cdots 
\f_{K_1} \rangle
  \quad.\label{pta}
\ee
In the case of field theories of particles, the
physically relevant quantity is (\ref{tra}), which gives the transition 
amplitude from a state in a distant (infinite) past to a state in a distant 
(infinite) future. 
Therefore in order to obtain
meaningful amplitudes one has to sum all Feynman diagrams up to a given order
of perturbation theory. However, in the case of field 
theories associated to simplical gravity, we will argue that even the 
individual
Feynman diagrams, or certain sub-classes of diagrams, can have a physical 
interpretation. This means that it is not necessary to sum over all possible
triangulations interpolating between the initial and the final surface.

Note for example that in the case of the vacuum diagrams,
a single diagram $\G$ is the partition 
function for the simplical complex $\ct$ where $\G = \ct^*_1$ is the dual 
one-complex to $\ct$. 
If one thinks about the simplical
gravity as the fundamental theory,i.e. not as an auxiliary device for
defining the path integral of the continuum theory, then each equivalence
class of simplical complexes could have a physical meaning. This is trivially
satisfied in the case of topological theories, where due to the 
triangulation independence, the continuum theory is the same as the
simplical theory, and hence the partition function for a single simplical 
complex $\ct (M)$ is the partition function for $M$.

\section{Partition function}

In order to develop these ideas further, we examine in some detail our
2d simplical gravity model. Consider the Mercedes-Benz (MB) diagram. It is
a dual one-complex for a tetrahedron $t$, if the $t$ is considered
as a 2d simplical complex. In that case $t=\ct (S^2)$, and hence
the value of the MB diagram will be the partition 
function for a sphere $S^2$. The kinetic term (\ref{tdk})
implies that the propagator will be given by two parallel lines, representing
the basic contraction
\be
\langle \f^\L_{\a\b}\,\f^{\pL}_{\pa\pb} \rangle = 
(-1)^{\L -\a -\b}d_\L \d^{\L,\pL} \d_{\a,-\pa}\d_{\b,-\pb}\,.\label{prop}
\ee
The interaction term (\ref{tdi}) implies that 
the vertex will be given by a triangle whose corners are cut, 
where the indices $\a_i \in \L_i$ are assigned to each edge. One then joins the
vertices by the propagator lines, and in this way one obtains a ribbon MB 
graph. The value of the diagram will be 
given by the product of the edge delta functions multiplied by the 
propagator factors and the vertex factors. In general there will be $2E$ 
delta functions, which will be multiplied
by a factor $(-1)^{E\L}(d_\L)^{E-2V}$, where $E$ is the number of edges 
and $V$ is the number of vertices of $\G$. For the MB case $E=6$ and $V=4$.
By summing over the indices along a loop of $\G$,
the corresponding delta functions give the factor  $d_\L$, so that 
\bea
Z (\G)  &=& \sum_\L (d_\L )^{E - 2V + F}(-1)^{E\L}\nonumber \\ 
&=& \sum_\L (d_\L )^{V - E + F }(-1)^{E\L} = 
\sum_\L ( d_\L )^{\c (M)}(-1)^{E\L} \quad,\label{tdss}  
\eea
where $F$ is the number of loops in the ribbon graph $\G$ ($F(MB)= 4$), 
which is the same as the number of faces in $\ct^*$,
and we have used that $2E=3V$ for trivalent graphs.

Since $\c (\G) = \c (\ct(M))= \c (M) = 2 - 2g$, the Euler number, is a 
topological 
invariant, where $g$ is the genus of $M$, $Z$ would be a topological invariant
provided that there is no sign factor $(-1)^{E\L}$. It comes from the
sign factor in the propagator (\ref{prop}). There are several ways 
one can get rid off this sign factor. One can 
change the kinetic term into
\be
{\tilde S}_k  
=\half \sum_{\L,\a,\b} d_\L^{-1} \f^\L_{\a\b} \f^\L_{-\a-\b} \,,\label{fmk}
\ee
so that the sign factor is removed from the propagator. Although this choice 
gives the correct answer, this action does not follow in general from the 
integration over the group, unless one has a group where one can choose a 
basis where $(D^\L_{\a\b})^* (g) = D^\L_{-\a-\b} (g)$. Similarly,
one can choose
\be
{\tilde S}_k
=\half \sum_{\L,\a,\b} d_\L^{-1} \left(\f^\L_{\a\b}\right)^2 \,, 
\label{smk}
\ee
and this would follow from the group integration if the group $G$ allows
a basis where the matrices $D(g)$ are real. In both cases
we will write the propagator as
\be
\langle \f (12) \, \f (34) \rangle = \d (13) \d (24) \quad. \label{unt} 
\ee

One can avoid the propagator factor $d_\L^{-1}$ by rescaling $\f$, which
then changes the vertex factor to $d_\L^{-\half}$. Either way
the propagator (\ref{unt}) gives the required result
\be
Z (M) = \sum_\L ( d_\L )^{\c (M)}\quad.
\ee

If one wants to have an integral over the group expression which is valid
for arbitrary $G$, then one can choose
\be
S_k =\half\int_{G^2}d^2 g\, \vf (g_1 ,g_2) \vf (g_2 ,g_1) 
=\half \sum_{\L,\a,\b} \f^\L_{\a\b} \f^\L_{\b\a} \,.
\ee
This choice gives a twisted propagator
\be
\langle \f (12) \, \f (34)\rangle = \d (14) \d (23) \quad. \label{tw}
\ee
The difference between the twisted propagator case (\ref{tw}) 
and the untwisted case (\ref{unt}) is that the same graph will not describe the
same surface in each case. For example, in the case of the theta graph,
the untwisted propagator gives $\c = 2$, a sphere, while the twisted 
propagator gives $\c = 0$, a torus. In the untwisted case the torus is obtained
by braiding two edges of the theta graph, which for the twisted case gives a 
sphere.

Note that one can impose the symmetry condition $\vf (1 ,2)= \vf (2 ,1)$,
which together with the reality condition implies
\be
(\f^\L_{\a\b})^* = \f^\L_{\b\a} \quad, 
\ee
i.e. $\f^\L$ are hermitian matrices. This gives a group theory generalization 
of the hermitian matrix models used in string theory.  

\section{Transition amplitudes and spin networks}

Now let us try to formulate a transition amplitude from a
spatial section $\S_1$ to a spatial section $\S_2$, which are boundaries of
$M$, i.e. we cut two holes in $M$. Consider the case when the 
corresponding simplical complex is a prism.  Then the base
triangles correspond to triangulations of $\S_i = S^1$, where 
$S^1$ is a circle. Each side of the 
prism is divided by a diagonal into two triangles, and this defines our 
triangulation $\ct (M)$ where $M = S^2$ a two-sphere. The corresponding 
Feynman diagram is given by the dual graph, which is a circle with six legs.
Its value can
be obtained by making appropriate contractions in
\be
\langle \f^{\L_1}_{\a_1 \b_1} \f^{\L_2}_{\a_2 \b_2} \f^{\L_3}_{\a_3 \b_3} 
 (S_v)^6 
\f^{\L_4}_{\a_4 \b_4} \f^{\L_5}_{\a_5 \b_5} \f^{\L_6}_{\a_6 \b_6} \rangle 
\label{slg}\quad.
\ee

If we take (\ref{slg}) as it is, we will obtain an expression which is not $G$ 
invariant, because of the uncontracted  representation indices. 
In order to 
remedy this, consider the following quantity, or the operator
\be
\F_3=\int_{G^3} d^3 g \,\vf (g_1 ,g_2) \vf (g_2 ,g_3) \vf (g_3 ,g_1) 
\quad. \label{sntwo}
\ee
It is proportional to $S_v$, 
so if we use (\ref{tdi}), we get
\be
{\F_3} = \sum_{\L,\a,\b,\g}d_\L^{-2}\f^\L_{\a\b}\f^\L_{\b\g}
\f^\L_{\g\a} = \sum_\L d_\L^{-2}\F_3 (\L) \quad.
\ee
We then propose that $\F_3 (\L)$ be an operator associated to a 
1d spin network given by a
triangle whose each side is colored by the irrep $\L$. Note that $\F_3(\L)$
is defined up to a scale factor, so we can write
\be
{\F_3}(\L ) = {1\over N(\L)} \sum_{\a,\b,\g}\f^\L_{\a\b}\f^\L_{\b\g} 
\f^\L_{\g\a} = {1\over N(\L)}\,Tr\, ( \f^\L )^3
\quad.
\ee
A natural normalization factor would be $N(\L) = d_\L^{3/2}$. 

One can then
associate states $\ket{\F_3(\L)}$ to operators $\F_3(\L)$, and the 
corresponding amplitude will be given by
\bea
A^{(6)}_{33} &=& \bra{\F_3 (\L)} (S_v)^6 \ket{\F_3 (\L)} = 
\langle \F_3 (\L) (S_v)^6 \F_3 (\L) \rangle \nonumber \\ 
&=& {1\over N^2 (\L)} \sum_{\a\b\g}
\langle \f^{\L}_{\a_1 \b_1} \f^{\L}_{\b_1 \g_1} \f^{\L}_{\g_1 \a_1} 
 (S_v)^6 
\f^{\L_4}_{\a_2 \b_2} \f^{\L}_{\b_2 \g_2} \f^{\L}_{\g_2 \a_2} \rangle 
\label{samp}
\quad.
\eea 

If we think of 
$\Psi_3$ as a quantity associated to a triangle whose each vertex carries a 
group
element $g$, then $\F_3 (\L)$ can be interpreted as a quantity corresponding
to a spin net formed by a spatially dual diagram to the triangle, where 
the duality is with respect to a 1d space, and hence
one obtains a triangle as a dual complex. Note that we can think of 
$\vf (g_1,g_2 )=\vf (1,2)$ as a quantity associated to the edge at whose 
ends are attached the group elements $g_1$ and $g_2$. Therefore the 
quantity
\be
\F_n  = \int_{G^n} d^n g \, \vf (1,2) \vf (2,3)\cdots \vf (n,1) 
\ee 
will correspond to an $n$-sided polygon, whose each vertex carries a group 
element. Hence $\F_n$ will
generate a spin network variable $\F_n (\L) \propto Tr\,(\f^\L)^n$, 
which will correspond to a closed 1d spin network with $n$ edges colored by 
$\L$. 

It is easy to see that
the non-zero transition amplitudes will come from the perturbative
amplitudes
\be
A^{(k)}_{lm} = \bra{\F_m (\L)} (S_v)^k \ket{\F_l (\L)} = 
\langle \F_m (\L) (S_v)^k \F_l (\L)\rangle \label{namp}
\quad.
\ee 

\section{Time variables}

Given this setup, and our remarks at the end of section 2, we can say 
something about possible time intervals for the amplitudes. The
amplitude for the prism (\ref{samp}) could be interpreted as a transition 
amplitude from
the state $\F_1$ to the state $\F_2$ in one unit of discrete time, so that 
the 
transition amplitude in $n$ units of time would be given by a sum of FD 
corresponding
to triangulations consisting of $n$ prisms put on 
top of each other. This suggests a discrete time variable $T$ to be
the minimal distance, i.e. the number of edges, from the initial to the final 
triangle in the corresponding triangulation. However, this time 
variable will make sense only for triangulations where the minimal 
distance between any two points of the initial and the final triangle
does not exceed $n+1$. In general case these are the triangulations
where for every point $p$ of the inital polygon $P$ we have
\be
 min\,\{ d(p,p^\prime )\,|\,p^\prime \in P^\prime \} = n \quad,\label{pod}
\ee
where $P^\prime$ is the final polygon. 

Note that the triangulations which
satisfy (\ref{pod}) can be sliced into $(n-1)$ polygons $P_k$, such that 
the distance (\ref{pod}) is one for each pair $(P_k,P_{k+1})$,
where $P_0 = P$ and $P_n = P^\prime$. It is easy to see that there will be
infinitely many such triangulations for every $n\ge 2$. Hence if
one tries to define the amplitude $A(n)$ for $n$ units of time as a sum over 
triangulations with $n-1$ slices, then for $n\ge 2$ the amplitude $A(n)$ 
will contain infinitely many FD. Therefore one will have the same problem as 
in the case of the unrestricted amplitude 
\be
A_{lm} = \langle \F_m (\L) \exp(i S_v) \F_l (\L)\rangle \label{tamp}
\quad.
\ee 

One can then try to restrict the infinite sum of FD in $A(n)$ by using the 
perturbation
theory, i.e. by taking only the FD up to a given order of $\l$. However,
one has to
insure that the corresponding amplitudes are consistent. Namely, the amplitudes
have to obey the composition rules, which come from the procedure of 
pasting two cylinders. This is equivalent to requiring that the amplitudes
are given by the matrix elements of an evolution operator $U(T)$, 
which satisfies
\be
U(T_1)U(T_2)= U(T_1 +T_2 ) \quad, \label{comp}
\ee
where $T$ is the evolution parameter, i.e. the time variable.

For example, in the case of particle field theories, one has
$U(T)=\exp(iS_v )=\exp(iTH_v)$, where the split $S_v = TH_v$ is possible
because of the fixed spacetime background structure. In the case of simplical
field theories, this split is not natural, and it is not clear how to do it.
One can then use the minimal distance time variable $T$, and the corresponding
$U(T)$ would be defined by requiring that its matrix elements are given by the
restricted amplitudes $A(n)$. However, it is not clear why would such an 
operator satisfy (\ref{comp}), and it is difficult to check this. 

On the other hand, note that $U(T)=(S_v)^T$  satisfies the 
composition law (\ref{comp}). Hence the amplitude (\ref{namp})
can be interpreted as the transition amplitude for $T=k$ units of time,
where the time variable is given by the number of vertices in the 
corresponding Feynman diagrams, or
equivalently, by the number of triangles in the corresponding triangulations.
Since the amplitude (\ref{namp}) is given by a finite sum of FD, 
the new evolution operator will give finite amplitudes for finite time 
intervals, provided that the individual FD are finite. This can be 
achieved by going to the quantum group formulation. 

Since this choice of time is related
to the covariant quantization procedure we are using, i.e. we are not splitting
the spacetime into space and time, as in the canonical quantization, we will 
call this evolution parameter the covariant time.

However, there is a 
peculiarity with the covariant evolution operator. Since it is natural to 
define 
$S_v$ to be a hermitian operator in the field theory Fock space, 
so that $e^{iS_v}$ is a unitary operator, then
$U(n)= (S_v)^n$ is  going to be only a hermitian operator, but not a 
unitary operator. This would mean a non-unitary time evolution of states. 
However, one has to keep in mind that we are not dealing here with the usual
domain of quantum mechanics, i.e. objects in a fixed background spacetime.
We have here states describing the whole universe,
which is in our toy model represented by a colored triangulation of $S^1$, 
and hence
one cannot apriori expect that the same rules apply as in the standard quantum 
mechanics. 

\section{D=3 model}

We now examine our constructions for the case of 3d gravity. 
The partition function for the 3d BF theory is generated by the FD of a
$\f^4$ field theory on $G^3$ \cite{b}, whose action is given by
\be
S_3 = \half\int_{G^3}d^3 g\, \vf^2 (1 ,2 ,3) + {\l\over 4!}\int_{G^6}
d^6 g \,\vf (1 ,2 ,3)\vf (1,4,5)\vf (2,5,6)\vf (3,6,4)
\quad,\label{thda}
\ee
where we use a short-hand $\vf (1,2,3) =\vf (g_1 , g_2 ,g_3)$ and $\vf$ 
satisfies
\be
\vf (g_1 , g_2 ,g_3) = \vf (g_1 g , g_2 g ,g_3 g )
\quad. \label{tri}
\ee
The condition (\ref{tri}) implies
\be
\vf (g_1 , g_2 ,g_3) = \int_G dg \,\f (g_1 g , g_2 g ,g_3 g )\label{intf} 
\quad,
\ee
where $\f$ is an unconstrained field. 

Note that the form of $S_3$ allows the following (category theory) 
interpretation. If we 
associate $\vf (1,2,3)$ to a triangle whose edges are labeled with the group 
elements $g_1, g_2 ,g_3$, the interaction term $S_v$ can be associated to a 
tetrahedron whose edges are labeled by the group elements $g_1,... ,g_6$,
and the triangles have orientations provided by the outer normals. 
With this interpretation it is natural to require the cyclic symmetry for 
$\vf$, i.e.
\be
\vf (1,2,3) = \vf (2,3,1) = \vf (3,1,2) \quad. 
\ee
However, as in the 2d case, one does not have to impose this symmetry.

The Fourier modes expansion of (\ref{intf}) gives
\be
\vf  = \sum_{\L,\a}\sqrt{d_{\L_1} d_{\L_2} d_{\L_3 }}
\,\f^{\L_1 \L_2 \L_3}_{\a_1 \a_2 \a_3  } 
D^{\L_1 \L_2 \L_3}_{\a_1 \a_2 \a_3 } (g_1 ,g_2 ,g_3) 
\quad,\label{fmod}
\ee
where
\bea
\f^{\L_1 \L_2 \L_3}_{\a_1 \a_2 \a_3 } &=& 
\sum_{\b} \f_{\a_1 \b_1 \a_2 \b_2 \a_3 \b_3}^{\L_1 \L_2 
\L_3} {\bar C}_{\b_1 \b_2 \b_3}^{\L_1 \L_2 \L_3}\,,
  \\ 
D^{\L_1 \L_2 \L_3}_{\a_1 \a_2 \a_3 } (g_1 ,g_2 ,g_3) &=&
\sum_{\b}D^{\L_1}_{\a_1 \b_1}(g_1) D^{\L_2}_{ \a_2 \b_2}(g_2) 
D_{\a_3 \b_3}^{\L_3}(g_3) 
C_{\b_1 \b_2 \b_3}^{\L_1 \L_2 \L_3}\,.
\eea
The coefficients $C$ are basis components of the intertwiner map $\i_3$
\be
\i_3 : V(\L_1) \otimes V(\L_2) \otimes V(\L_3) 
\to V(0) \quad,
\ee
where $V(0)$ is the subspace of the singlets, so that 
\be
\ket{0;\L_1 ,\L_2 ,\L_3 }= \sum_\a C_{\a_1 \a_2 \a_3}^{\L_1 \L_2 \L_3} 
\ket{\a_1}\otimes\ket{\a_2}\otimes\ket{\a_3}\,,
\label{cbas}
\ee
and $\ket{\a}$ are vectors of an orthonormal basis.

In order to obtain (\ref{fmod}) one needs the relation
\be
\int_G dg \,U^{\L_1}(g) \otimes U^{\L_2}(g) \otimes 
U^{\L_3}(g) = \i_3 {\i_3}^\dagger \quad, \label{proj}
\ee
which in the basis (\ref{cbas}) becomes
\be
\int_G dg \,  D^{\L_1}_{\a_1 \b_1} (g) D^{\L_2}_{\a_2 \b_2} (g) 
D^{\L_3}_{\a_3 \b_3} (g) 
= 
C^{\L_1 \L_2 \L_3}_{\a_1 \a_2 \a_3 }{\bar C}^{\L_1 \L_2 \L_3}_{\b_1 \b_2 \b_3 }
\quad. 
\ee

By inserting (\ref{fmod}) into $S_3$ we obtain
\be
S_k = \half \sum_{\L,\a} \left| \f^{\L(123)}_{\a(123)}\right|^2 \,,\label{thka}
\ee
and
\be
S_v = {\l\over 4!}\sum_{\L ,\a } (-1)^{\sum_i (j_i - \a_i)}
\f^{\L(123)}_{\a(123)}\,\f^{\L(145)}_{\a(-145)}\,\f^{\L(256)}_{\a(-2-56)}
\,\f^{\L(364)}_{\a(-3-6-4)}
\left\{ \begin{array}{ccc}\L_1 & \L_2 & \L_3 \\ 
\L_4 & \L_5 & \L_6 \end{array} \right\}
 \,,\label{thia}
\ee
where we have used a shorthand $X(jkl)= X_j X_k X_l$ and 
\be
\left\{ \begin{array}{ccc}\L_1 & \L_2 & \L_3 \\ 
\L_4 & \L_5 & \L_6 \end{array} \right\} = \sum_{\b} (-1)^{\sum_i (j_i - \b_i)}
C^{\L(123)}_{\b(123)}\,C^{\L(145)}_{\b(-145)}\,C^{\L(256)}_{\b(-2-56)}
\,C^{\L(364)}_{\b(-3-6-4)}
\ee
is the $6j$ symbol. We also denote it as $6j(\L_1 \cdots \L_6 )$. 

Since the reality 
condition can be written as
\be
\left(\f^{\L(123)}_{\a(123)}\right)^* = (-1)^{\sum_i (j_i - \a_i)}
\f^{\L(123)}_{\a(-1-2-3)}
\,,
\ee
then the kinetic term (\ref{thka}) gives for the propagator 
\be
\langle\f^{\L(123)}_{\a(123)}\, \f^{\pL(123)}_{\pa(123)}\rangle 
=\prod_{i=1}^3 (-1)^{j_i - \a_i} \d^{\L_i , \pL_i}
\d_{\a_i ,-\pa_i}
\quad. \label{thdp} 
\ee
Note that the 3d propagator contains a sign factor analogous to the one in the 
2d case. Since for our purposes this sign factor, as well as the related
sign factors, will not be important, we will omit them from the formulas. We
will then write the formulas as for the real $D(g)$ case.

The Feynman diagrams will be given by the four-valent graphs $\G$, where
each edge is represented by 3 parallel lines, while each vertex is represented
by a tetrahedron whose corners are cut and the corresponding edges are joined 
with the propagator lines. The Feynman rules which follow from (\ref{thka})
and (\ref{thia}) consist of
assigning an index $\a_i \in \L_i$ to each  
edge of each tetrahedron, as well as the corresponding $6j$ symbol.
One then assigns a delta function for each propagator line and 
multiplies all the delta
functions with the $6j$ symbols. If $\G$ has $|V|$ vertices and $|E|$ edges,
one then obtains a product of $|V|$ $6j$ symbols and $3|E|$ delta functions.
By summing over the indices for each loop\footnote{This is also a face if $\G$ is considered as a two-complex. This two-complex is called a spin foam
\cite{baez}.} $F$ of $\G$ one obtains a 
$d(\L_F)$ factor, so that the value of the diagram is given by
\be
 Z (\G ) = \sum_{\L_F} \prod_{F\in \G}\,d(\L _F ) 
\prod_{V\in \G} \, 6j\,(\L_{F_1(V)} \cdots \L_{F_6 (V)}) \,.
\ee

For example, 
consider a vacuum diagram given by the pentagram graph.
The pentagram is a dual one-complex
to the 4-simplex if the 4-simplex is considered as a 3d simplical complex.
This 4-simplex can be considered as  
a triangulation of the spacetime manifold $S^3$. 

If the graph $\G$ is dual to a triangulation $\ct (M)$, we can rewrite $Z$ as
\be
Z(M)= \sum_{\L_e} \prod_{e\in \ct}\,d(\L_e)\, \prod_{t\in \ct} \,
6j\,(\L_{e_1 (t)} \cdots \L_{e_6 (t)})
\ee 
where $e$ are the edges of $\ct$, which correspond to the faces $F$ of 
$\G$. $t$ are the tetrahedrons of $\ct$, which correspond to the vertices $V$
of $\G$. In this way one reproduces the Ponzano-Regge formula, where
the edges $e$ of the triangulation $\ct$ are colored with the $SU(2)$ irreps, 
and the $6j$ symbols are assigned to the tetrahedrons $t$. 
The sum over the colorings is divergent, and it can be regularized by using
the quantum group $SU_q (2)$ \cite{tv}. The result 
is triangulation independent, so that $Z_q$ is a topological invariant of $M$.

As far as the transition amplitudes are concerned, they will be given by a 
straightforward generalization of the $D=2$ amplitudes. For example,
if we wish to calculate the transition amplitude from an initial surface 
$\S_{1} = S^2$ to a final surface $\S_{2} = S^2$, we take a 3d simplical 
complex with 
two disjoint boundary tetrahedrons $t_i$, which are taken to be the 
triangulations of the 
initial and the final surface. The corresponding Feynman diagram will be 
given by 
the dual one-complex, i.e. a four-valent graph with 8 external legs.
For the invariant amplitude one has to consider the $D=3$ analog of
(\ref{sntwo}), i.e. to a tetrahedron whose edges are labeled with group 
elements we assign
\be
\F_4 =\int_{G^6} d^6 g \,\vf (1,2,3) \vf (1,4,5) \vf (2,5,6)\vf (3,6,4)
\quad. \label{sntetr}
\ee

By expanding $\vf$'s in the Fourier modes we obtain
\be
\F_4  = \sum_{\L,\a }
\f^{\L(123)}_{\a(123)}\,\f^{\L(145)}_{\a(145)}\,\f^{\L(256)}_{\a(256)}
\,\f^{\L(364)}_{\a(364)}
\left\{ \begin{array}{ccc}\L_1 & \L_2 & \L_3 \\ 
\L_4 & \L_5 & \L_6 \end{array} \right\}\,.
\ee
Hence we define
\be 
\F_4 (\L_1,\cdots,\L_6)= {1\over N(\L)}\sum_{\a }
\f^{\L(123)}_{\a(123)}\,\f^{\L(145)}_{\a(145)}\,\f^{\L(256)}_{\a(256)}
\,\f^{\L(364)}_{\a(364)}\quad,
\ee
where a natural normalization would be $N(\L)=\sqrt{d(\L_1)\cdots d(\L_6)}$.

We then associate the quantity $\F (\L_1,\cdots,\L_6)$ to the 
spin net corresponding to a
triangulation of $S^2$ given by a tetrahedron $t$. The spin net graph will be
given by the dual one-complex
to  $t$, where $t$ is considered as a 2d simplical complex. One then obtains
the Mercedes-Benz graph, and its edges will be labeled with the irreps 
$\L_i$. Hence a perturbative amplitude for the transition from 
$\ket{\F(\L_1 ,\cdots, \L_6)}$ to $\ket{\F(\pL_1 ,\cdots ,\pL_6)}$ will be 
given by
\be
A_{if}(n) = \langle \F (\pL_1,\cdots\pL_6 ) (S_v)^n \F (\L_1,\cdots ,\L_6)
\rangle 
\,, \label{thdta}
\ee 
where $n\ge 8$.

If we consider a $\ct (S^2)$ which is two tetrahedrons with a common face, then
the dual graph is a  prism. If we attach to the edges of $\ct$ the 
group elements $g_1,\cdots , g_9$, we can then construct an invariant operator
\be
\Psi_6 =\int_{G^9} d^9 g \,\vf (1,4,2) \vf (2,5,3) \vf (1,3,6)\vf (4,7,8) 
\vf (5,8,9) \vf (7,6,9)
\quad. \label{snprism}
\ee 
By expanding into the Fourier modes we obtain 
$\F_6 (\L_1 \cdots \L_9)$, which is an operator associated
to the spin net given by a prism whose edges are labeled by the
irreps $\L_1 \cdots \L_9$. In this way we obtain operators and states for  
2d spin nets whose graphs are dual to triangulations of a boundary
surface $\S$. This surface represents a space part of the spacetime manifold 
$M$. 

Therefore we denote the $\F_n (\L)$ operators and states as
$\F (\g ,\L)$, where $\g$ denotes the spin net graph, which is dual to 
$\ct(\S)$, and $\L$ denotes the labeling of the edges of $\g$ with the unitary
irreps of $G$. One can write
\be 
\F (\g ,\L )= {1\over N(\L)}\sum_{\a }\prod_{v}
\f^{\L (e_1 (v) e_2 (v) e_3 (v))}_{\a (e_1(v) e_2 (v) e_3 (v))}
\quad, \label{thdsnf}
\ee 
where $v$ are the vertices of $\g$, and $e_i (v)$ are the three edges coming
out of the vertex $v$.

The amplitude (\ref{thdta}) is equal to a sum of FD with $n$ 
vertices and $6+6$ external legs. The value of the each diagram in that sum
can be written as
\be
A(\ct) =N(\ct) \sum_{\L_{\tilde e}}6j (\pL_1 \cdots\pL_6 )\prod_{\tilde e \in\ct}
d(\L_{\tilde e})  \prod_{\tilde t \in \ct}
6j(\L_{\tilde e_1 (\tilde t)} \cdots \L_{\tilde e_6 (\tilde t)})
 6j(\L_1 \cdots \L_6 ) 
\,, \label{prta}
\ee 
where 
\be
N(\ct) = {d(\pL_1)\cdots d(\pL_6) d(\L_1)\cdots d(\L_6)\over 
N(\pL_1 \cdots \pL_6 ) N(\L_1 \cdots \L_6 )}=
\sqrt{d(\pL_1)\cdots d(\pL_6) d(\L_1)\cdots d(\L_6)}\,,
\ee
and $\tilde t$ are the interior tetrahedra and $\tilde e$ are
the interior edges of a triangulation $\ct$ which is dual to the Feynman 
diagram, 
while the  edges of the boundary
tetrahedra are labeled with the irreps $\L_1,...,\L_6$ and $\pL_1,...,\pL_6$.

The amplitude $A(\ct)$ can be recognized as the Ponzano-Regge amplitude for 
the transition
from an initial surface to a final one via triangulation $\ct$ of the 
spacetime manifold whose boundaries are the initial and the final surface,
where the edges of the initial and the final triangulations are colored with 
a fixed set of irreps \cite{por,fr}. 

The general spin net amplitude
\be
A_{if}(n) = \bra{\F(\g^\prime,\pL)} (S_v)^n \ket{\F(\g ,\L)} 
\,, \label{sntda}
\ee 
can be computed from
\be
A_{if}(n) = \langle\F(\g^\prime,\pL) (S_v)^n \F(\g ,\L) \rangle 
\,,
\ee 
by using the formula (\ref{thdsnf}). It
will be given by the sum of all $(v,v^\prime )$ FD with $n$ vertices and 
$v + v^\prime$ external legs, multiplied by the normalization factor
$N^{-1}(\L)N^{-1}(\pL)$, where $v$ and $v^\prime$ are the numbers of 
vertices in the spin nets $\g$ and $\g^\prime$, respectively. After summing over
the representation indices, each $FD$ will give the Ponzano-Regge amplitude 
for the corresponding 3-complex $\ct$ consisting of $n$ tetrahedrons and 
with two disjoint boundaries consisting of $v$ and $v^\prime$ tetrahedrons. 
One can then write
\be
A_{if}(n) = \sum_{\ct,|\ct|=n} A(\ct ;\g,\L ;\g^\prime,\L^\prime ) \,.
\label{prna}
\ee
The 
colored boundary complexes $\ct_v$ and $\ct_v^\prime$ are dual to the spin nets
$\g$ and $\g^\prime$ since they can be considered as two-complexes 
corresponding to triangulations of the initial and the final surface. 

Also, according to our time variable interpretation, the amplitude 
(\ref{prna}) will correspond
to the transition from the initial to the final state in $n$ units of 
the covariant time.   

Note that the spin network states $\ket{\F(\g ,\L)}$ have the same labeling
as the spin network states $\ket{\Psi(\g ,\L)}$ obtained by the canonical 
quantization of 3d gravity, or equivalently, the $SU(2)$ BF theory 
\cite{baez}. However,
these states are fundamentally different, because they belong to different
Hilbert states, which is the consequence of two different quantization 
procedures which are used. 

In the canonical approach, the spacetime manifold $M$ has a fixed topology
${\bf R}\times \S$, where $\S$ is a 2d surface. In the connection 
representation, one can construct a Hilbert space of gauge invariant states
$L_2 (\ca / \cg )$, whose orthonormal basis is given by the spin network 
states $\Psi_{\g ,\L} (A)$. These are evaluated by assigning the 
open holonomies $U^{\L}(e,A)$ to the edges $e$ of the spin net graph $\g$ which 
is embedded in 
$\S$, where $\L$ is the irrep assigned to the edge $e$. One then takes the
product of the holonomy matrices and contracts the indices with the 
intertwiners in order to obtain a scalar wave-function $\Psi_{\g,\L}(A)$.

This construction can be discretized \cite{fr,baez}, where $\S$ is replaced 
by a triangulation $\ct (\S)$, and $\g$ is the dual graph, while the
connection is replaced by a set of group elements $g_e$ representing the
holonomies of the edges $e$, so that 
$U^\L_{\a\b} (e,A) \to D^\L_{\a\b} (g_e)$. Then $\Psi$ 
becomes an element of the space $L_2 (G^E)$ where $E$ is the number of the 
edges of $\g$ \cite{al}.
Hence in this formulation, one can consider only the spin nets with a fixed
number of the edges, so that one cannot calculate the transition amplitude
between the spin nets having different numbers of edges. On the other hand, 
in the
case of field theory spin networks, there is no such a constraint, because
the Hilbert space has the Fock space structure, and one can have spin net
states with different numbers of the edges. Hence the field theory Hilbert
space contains states describing all colored 2-complexes, or equivalently,
the states for  manifolds $\S$ of all topologies.

Note that a single complex amplitude in (\ref{prna}) can be written as a 
scalar product
of the canonical quantization spin net states $\ket{\Psi(\g,\L)}$ as
\be
A(\ct,\g,\L ;\g^\prime ,\pL ) = \langle\Psi(\g^\prime,\pL)|\Psi(\g,\L)\rangle
\,,\ee
when $\g$ and $\g^\prime$ have the same numbers of edges \cite{o2,r2}. 
On the other hand, it follows from (\ref{prna}) that
\bea
A_{if}(n)&=& C(n, \g , \g^\prime )\langle\Psi(\g^\prime,\pL)|\Psi(\g,\L)\rangle
\nonumber \\
&=& \bra{\F(\g^\prime,\pL)} (S_v)^n \ket{\F(\g ,\L)}
\,,\label{cqsnta}
\eea  
where $C(n,\g,\g^\prime)$ is the number of 3-complexes $\ct$ with $n$ 
tetrahedrons
whose boundary complexes are dual to $\g$ and $\g^\prime$, or equivalently,
it is the number of $(v,v^\prime)$ FD with $n$ vertices. 
Hence due to the topological nature of 3d gravity, one can relate
the canonical and the field theory spin net
states.

\section{D=4 model}

The case of four spacetime dimensions is special for the BF theory, since
this is the dimension where the BF theory starts to be different from the GR 
theory. GR in $D\ge 4$ is not a topological theory, or in other words,
it has local degrees of freedom. This is also reflected by the
fact that the two-form $B_{ab}$, where $B=B^{ab}J_{ab}$ and $J_{ab}$ are the 
$SO(4)$ or $SO(3,1)$ Lie algebra generators, cannot be always written as 
$e^a \wedge e^b$, where $e^a$ is the fierbein one-form. Recall that in 3d 
$B^{ab}=\e^{abc}e_c$, so that the BF action becomes identical to the Palatini 
form of the Einstein-Hilbert action. Therefore in order to obtain a simplical 
non-topological gravity theory from the BF theory,
something has to be modified
in a 4d analog of the constructions described in the previous sections.

The field theory which generates the partition function of the topological
gravity theory in 4d was worked out by Ooguri \cite{o}. The labeling
pattern in the 4d action is a 
straightforward
generalization of the $D=2,3$ theories. We assign $\vf (g_1,...,g_4)$
to a tetrahedron whose faces are labeled with the group elements $g_i$, 
and the interaction term is given by a 4-simplex built out of 5 tetrahedrons,
so that
\bea
&S_4 & = \half\int_{G^4}d^4 g\, \vf^2 (1 ,2 ,3,4) \label{fda}\\
 &+& {\l\over 5!}\int_{G^{10}}
d^{10} g \,\vf (1 ,2 ,3,4)\vf (4,5,6,7)\vf (7,3,8,9)\vf (9,6,2,10)
\vf (10,8,5,1). \nonumber
\eea

Since $\vf$ has to be $G$-invariant, we have
\be
\vf (g_1 , g_2 ,g_3,g_4) = \int_G dg \,\f (g_1 g , g_2 g ,g_3 g , g_4 g )
\equiv P_G \f (g_1  , g_2  ,g_3  , g_4  )
\label{ginv} 
\quad.
\ee
This gives the following momentum mode expansion
\be
\vf  = \sum_{\L,\pL,\a}\sqrt{d(\L_1 ) d(\L_2 ) d(\L_3 ) d(\L_4 )}\,
\f^{\L(1234)\pL}_{\a(1234)} 
D^{\L(1234)\pL}_{\a(1234)} (g_1 ,g_2 ,g_3 ,g_4) 
\quad,\label{mom}
\ee
where
\be
\f^{\L(1234) \pL}_{\a(1234)} = 
\sum_{\b} 
\f_{\a\b (1234)}^{\L (1234)}\, {\bar C}_{\b (1234)}^{\L(1234) \pL}\,,
\ee
and
\be
D^{\L(1234) \pL}_{\a(1234) } (g_1 ,g_2 ,g_3,g_4) =
\sum_{\b}\prod_{i=1}^4 D^{\L_i}_{\a_i \b_i}(g_i) 
C_{\b(1234)}^{\L(1234) \pL}\quad.
\ee
The coefficients $C$ are components of the intertwinwer map $\i_4$
\be
\i_4 : V(\L_1) \otimes V(\L_2) \otimes V(\L_3) \otimes V(\L_4) 
\to V(0) \quad,
\ee
in an orthonormal basis.

In order to obtain (\ref{mom}) one needs the relation
\be
\int_G dg \,U^{\L_1}(g) \otimes U^{\L_2}(g) \otimes U^{\L_3}(g) 
\otimes U^{\L_4}(g) 
= \i_4 {\i_4}^\dagger \quad, \label{fdpro}
\ee
which in an orthonormal basis $\ket{\a}$ becomes
\be
\int_G dg \,  D^{\L_1}_{\a_1 \b_1} (g) D^{\L_2}_{\a_2 \b_2} (g) 
D^{\L_3}_{\a_3 \b_3} (g) D^{\L_4}_{\a_4 \b_4} (g) 
= \sum_{\pL}
C^{\L(1234) \pL}_{\a(1234) } C^{\L(1234) \pL}_{\b(1234) }
\quad, 
\ee
where $\pL$ labels the singlets in the tensor product of four irreps 
$\L_1,...,\L_4$.

By inserting (\ref{mom}) into $S_4$ we obtain for the kinetic part
\be
S_{k} = \half \sum_{\L,\pL,\a} \left| \f^{\L(1234)\pL}_{\a(1234)}\right|^2 
\quad, \label{fdk}
\ee
while for the interaction part we get
\be
S_{v}= {\l\over 5!}\sum_{\L ,\pL ,\a }
\f^{\L(1234)\pL_1}_{\a(1234)}\,\f^{\L(4567)\pL_2}_{\a(4567)}\,
\f^{\L(7389)\pL_3}_{\a(7389)}
\,\f^{\L(96210)\pL_4}_{\a(96210)}\f^{\L(10851)\pL_5}_{\a(10851)}
\left\{\begin{array}{ccc}\L_1 &\cdots& \L_{10}\\ \pL_1 &\cdots& \pL_5 
\end{array}\right\},\label{mma}
\ee 
where the $15j$ symbol is defined as
\be
\left\{\begin{array}{ccc}\L_1 &\cdots& \L_{10}\\ \pL_1 &\cdots& \pL_5 
\end{array}\right\}
= \sum_{\a} 
C^{\L(1234)\pL_1}_{\a(1234)}\,C^{\L(4567)\pL_2}_{\a(4567)}\,
C^{\L(7389)\pL_3}_{\a(7389)}
\,C^{\L(96210)\pL_4}_{\a(96210)}\,C^{\L(10851)\pL_5}_{\a(10851)}.\label{fjs}
\ee
We have written the formulas (\ref{mma}) and (\ref{fjs}) for 
real $D(g)$ matrices. In the complex case 
the repeated $\a$ indices will have a minus sign and there will
be the sign factors coming from the complex conjugation rules for
the $D (g)$ matrix elements. Since this will not be important for our 
purposes, we will use the simpler real basis notation.

The kinetic action (\ref{fdk}) implies for the propagator
\be
\langle\f^{\L(1234)\L}_{\a(1234)}\, \f^{\pL(1234)\pL}_{\pa(1234)}\rangle 
=\d^{\L , \pL}\prod_{i=1}^4 \d^{\L_i , \pL_i}\d_{\a_i ,\pa_i}
\quad. \label{fdp} 
\ee
Note that for a tetrahedron with labeled faces it is natural to consider 
$\vf$ which is 
invariant under the permutations of the labels. This would then give the
constraints on the Fourier modes (the analog of the hermiticity in 2d case),
which would involve the $6j$ symbols \cite{o}. As a result, the propagator 
will not have the delta function for the intertwiner labels, but it will 
have the $6j$ symbol for $\L_i,\L$ and $\pL $. Since it is not clear whether 
the 
symmetric $\vf$ is necessary, and in order to make the presentation simpler,
we will not impose the permutation symmetry.

The Feynman diagrams will be given by the 5-valent graphs. These can be 
represented as ribbon graphs,
where each edge is replaced with four parallel lines, and each vertex is
replaced with a pentagram whose corners are cut, and the incident lines for
a corner are joined with the  four
lines of an edge. The value of a FD can be obtained by coloring the ten lines
of each pentagram with ten irreps $\L$, as well as
every corner with the intertwiner label $\pL$. One then assigns to each
pentagram the 
corresponding $15j$ symbol. For each propagator line one takes a delta function
of the labels it connects. One then takes the product of all delta 
functions with the $15j$ symbols, and sums over the labels.
 
Consider a vacuum FD $\G$ with $|V|$ vertices, which is a dual one complex
to a 4d simplical complex $\ct$ consisting of $|V|$ 4-simplices. For example, 
the hexagon
graph, a six vertex 5-valent graph with all the vertices connected, 
is a dual one-complex for a 5-simplex considered as a 4d simplical
complex. This 4d complex can be considered as a triangulation $\ct$ of 
$S^4$, which
consists of five 4-simplices. Since each
closed propagator line gives a $d(\L)$ factor, the Feynman rules associate to 
each loop (face) $F$ of $\G$ 
an irrep $\L_F$ with the weight $d(\L_F)$. For each
vertex there is the 15j symbol of 10 loops (faces) which share it plus 5 
intertwiner labels of the 5 edges coming out of the vertex. Hence one obtains 
\be
Z (\G) = \sum_{\L_F,\pL_E} \prod_{F\in \G}\,d(\L_F) 
\prod_{V\in \G} \, \left\{\begin{array}{ccc}\L_{F_1(V)} &\cdots& 
\L_{F_{10}(V)}\\ 
\pL_{E_1(V)} &\cdots& \pL_{E_5 (V)}\end{array}\right\} \,. \label{vfd}
\ee

Since a triangle $f$ of $\ct$ is dual to a face $F$ of $\G$, a tetrahedron 
$t$ to the edge $E$, and a four-simplex $s$ to a vertex $V$, we can rewrite 
the above expression as
\be
Z(M) = \sum_{\L_f ,\pL_t} \prod_{f\in \ct}\,d(\L_f)\, 
\prod_{s\in \ct} \,
\left\{\begin{array}{ccc}\L_{f_1 (s)}&\cdots& \L_{f_{10} (s)} \\
\pL_{t_1 (s)} &\cdots& \pL_{t_{5} (s)}\end{array}\right\}\,.\label{ss}
\ee
Hence $Z(M)$ is a sum over colorings of the complex $\ct(M)$, where each
triangle $f$ of $\ct$ is colored with an irrep $\L_f$, while each tetrahedron
$t$ of $\ct$ is colored with an intertwiner label $\pL_t$ from the tensor 
product of four irreps coloring the triangles of $t$.
This sum is divergent, and by going to the quantum group formalism
one obtains a finite number
which is independent of the triangulation, and hence represents a 
topological invariant of the 4-manifold \cite{cy}. 

The form (\ref{ss}) of the
partition function is the state sum which mathematicians have explored 
over
the years, and it was in this approach that the modification which lead to 
quantum GR was discovered \cite{bce}. Namely, the constraint 
$B^{ab}= e^a \wedge e^b$, which transforms the BF action into Pallatini 
action, can be written in a simple form as
\be
B^{ab}\wedge B_{ab} = 0 \label{bcon}\quad,
\ee 
where the indices are contracted with the group metric. The constraint 
(\ref{bcon}) is translated
in the state-sum formalism into a constraint on the representations with 
which one colors the triangles of a triangulation as
\be
J^{ab} J_{ab} = 0 \label{rcon}\quad,
\ee 
where $J^{ab}$ are the generators of the $SO(4)$ Lie algebra $so(4)$. 
Since $so(4)= su(2)\oplus su(2)$, the constraint (\ref{rcon}) becomes
\be
{\vec J}^2 - {\vec K}^2 = 0 \quad,\label{eucl}
\ee
where ${\vec J}^2$ and ${\vec K}^2$ are the Casimirs of the two $SU(2)$. 

In terms of 
representations this constraint implies that one labels the triangles $f$ of a
4-simplex $s$ with the simple irreps $N=(j,j)$, and
instead of summing over all
irreps $\L = (j,k)$ in the state sum,
one sums only over the simple irreps. Then the $15j$ symbol
becomes a sort of $10j$ symbol, the Barret-Crane vertex
\be
\cv (N_1 \cdots N_{10})
= \sum_{M,\a} 
C^{N(1234)M_1}_{\a(1234)}\,C^{N(4567)M_2}_{\a(4567)}\,
C^{N(7389)M_3}_{\a(7389)}
\,C^{N(96210)M_4}_{\a(96210)}\,C^{N(10851)M_5}_{\a(10851)} \quad,
\ee
where $M$ labels the simple irrep singlets appearing in the tensor product of
four simple irreps. Hence
\be
Z_{BC}(\ct) = \sum_{N_f} \prod_{f\in \ct}\,d(N_f)\, 
\prod_{s\in \ct} \,
\cv (N_{f_1 (s)} \cdots N_{f_{10} (s)})
\,,\label{bcss}
\ee
where we have written $Z$ as a function of $\ct$, since now the state sum will
depend on the local parameters of the triangulation $\ct$.

An important property of the BC vertex is that it can be expressed in terms 
of an integral over $SU(2)^5$ \cite{barr}
\be
\cv (N_1 \cdots N_{10})
= \int_{SU(2)^{5}} d^5 h \prod_{1\le k < l \le 5} Tr_{j_{kl}}(h_k h_l^{-1}) 
\,,\label{vf}
\ee
where $(j_{12},...,j_{45}) = (N_1,...,N_{10})$ label the edges of a 
4-simplex $s$,
and $h_i$ label the vertices. This formula can be easily proved by 
noticing that the right-hand expression is $[10j(SU(2))]^2$ and since 
$SO(4) \simeq SU(2)\times SU(2)$, then 
$10j(SO(4))|_{j=simple} = [10j(SU(2))]^2$.

The explanation of this
formula comes from the fact that 
$$ SU(2)\simeq S^3 = SO(4)/SO(3)\quad, $$ 
so that the
$h$ variables are really coordinates $x$ on the homogeneous space $X=G/H$, 
which is a consequence of the fact that $N=(j,j)$
is a class-one representation of $G=SO(4)$ with respect
to the subgroup $H=SO(3)$ \cite{fk}. The quantity $Tr_j (h_k h_l^{-1})$ is the 
zonal 
spherical function, and can be thought of as a Green's function on the $X$ 
space, so that the BC vertex can be represented as a Feynman graph of a field
theory which lives on $X$. One then obtains
\bea
G_N (x_1,x_2) &=& Tr_j (h_1 h_2^{-1})\label{trf}\\
&=& {\sin ((2j + 1)\th_{12})\over
\sin \th_{12}}\quad, \label{szf}
\eea
where $\th_{12}$ is the angle between the unit 
4-vectors $x_i$, or equivalently, the geodesic distance between the 
points of $X$. The vectors $x_i$
can be interpreted as the normals to the tetrahedrons $t_i$ which share
the face $f_{12}$ labeled with the irrep $j$. 

The representation (\ref{vf}) combined with the formula (\ref{szf}) gives
the asymptotics of $\cv$ in the limit of large $j$ 
\be
\cv \sim  \sum_{\ca} \D (A)\cos\left(\sum_{f\in s} A_{f}\th_{f}+
\p /4 \right)\,, \label{bcva}
\ee
where $A_f = 2j_f +1$ is the area of the face $f$, and one has to sum over
certain inequivalent colorings $\ca$ of the four-simplex $s$, for details see 
\cite{bw}. When (\ref{bcva}) is inserted in $Z_{BC}$, it is easy to see that 
the terms of the
form $Q e^{iS_R} $ will appear, where $S_R = \sum_{f\in \ct} A_{f}\d_{f}$
is the Regge form of the EH action. This is an indication that the 
semi-classical 
limit of the BC model is a theory that is related to classical GR. 
However, the semi-classical limit of the BC model  
has to be investigated in more detail in order to be able to make more definite
statements.

Another important consequence of the formula (\ref{vf}) is that it can be 
easily 
generalized to the Lorentzian case \cite{bcl}, since then 
$G=SO(3,1)\simeq SL(2,C)$, 
$H=SO(3)$ and therefore $X$ is a hyperboloid. The unitary irreps of $SL(2,C)$
are labeled with two numbers, i.e. $\L = (n,p)$ where $n\in {\bf Z}$ and 
$p\in {\bf R}$,
and the simple representations are $(0,p)$ and $(n,0)$. It is natural to assign
$(0,p)$ irreps to space-like triangles and $(n,0)$ to  time-like 
triangles. One can restrict the model to the continuous irreps only 
so that 
\be
G_p (x_1 ,x_2 ) = {\sin (p\, d_{12})\over p\sinh d_{12}} \quad,\label{lgf}
\ee
where $d_{12}$ is the geodesic distance between $x_1$ and $x_2$. By using
(\ref{lgf}) one can construct the Lorentzian analog of the BC vertex via 
\be
\cv (N_1 \cdots N_{10})
= \int_{X^5} d^5 x \prod_{1\le k < l\le 5} G_{p_{kl}}(x_k , x_l) \quad, 
\label{lv}
\ee
where $(N_1 ,...,N_{10}) = (p_{12},...,p_{45})$.

As discussed in \cite{fk} for the Euclidean case and in \cite{bcl} for the
Lorentzian case, the constructions (\ref{vf}) and (\ref{lv}) can be 
generalized to give expressions for higher $nj$ symbols. Given a  
graph $\g$ we can label the vertices of $\g$ with the coordinates $x_k$
from the homogeneous space $X=G/H$. Then associate to each edge $(k,l)$ of 
$\g$ a simple irrep labeled with $j_{kl}$. In this way one obtains a labeled
graph which is called a relativistic spin net. We can than
associate the ``propagator'' $G_{j_{kl}} (x_k , x_l )$ to each edge,
and if we now consider $\g$ as a Feynman diagram of a theory with delta
function vertices, then its value will be given by 
\be
\cv (N_1 \cdots N_{n})
= \int_{X^m} d^m x \,\prod_{1\le k < l \le m}\, G_{j_{kl}}(x_k , x_l) 
\quad, \label{snv}
\ee
where
$\{ j_{kl} |1\le k<l \le m \}=\{N_1 ,\cdots, N_{n} \}$.  

In the Euclidean case, the expression (\ref{snv}) is finite, since it is 
equal to a finite sum of  products of a finite number of $C$ symbols. 
This is a consequence of the formula (\ref{trf}).
In the Lorentzian case, the finiteness is not obvious. The $10j$ symbol was 
conjectured to be finite in \cite{bcl}, and recently a proof of finiteness
was given in \cite{babar}. 

As far as the finiteness of the state sum $Z_{BC} (\ct)$ is concerned,
it is divergent, but as argued in \cite{bcl}, one expects to become finite
in the quantum group formalism. It is interesting that there is a 
modification of the
BC field theory model in the Euclidean case which seems to give finite $Z$
for all $\ct$ \cite{pr,per}. 

\section{BC field theory}

Now we can discuss the field theory formulation of the BC model \cite{dfkr}.
The constraint of using only the simple representations to label the triangles
of a triangulation is translated in the field theory formalism into
a requirement of invariance under the $SO(3)$ subgroup: 
$\vf (g_i) = \vf (g_i h_i)\,,\, h_i \in H$. Hence
\be
\vf (g_1 , g_2 ,g_3,g_4) = \int_{H^4} d^4 h \,
\f (g_1 h_1 , g_2 h_2 ,g_3 h_3 , g_4 h_4  )\equiv P_H \f\label{hin} 
\quad,
\ee
where $\f$ is an unconstrained field.
By combining the $H$-invariance (\ref{hin}) with the usual $G$-invariance 
(\ref{ginv}) one obtains
\bea
\vf &=& P_G P_H \f = \int_G dg \int_{H^4} d^4 h \,\f (g_i g h_i )\nonumber \\
&=& \sum_{N,\a,\b} \sqrt{d(N_1)\cdots d(N_4)} \f^{N(1234)}_{\a(1234)}
\prod_{i=1}^4 D^{N_i}_{\a_i \b_i} (g_i) B^{N(1234)}_{\b(1234)}\,, \label{ghfe}
\eea
where 
\be
B^{N(1234)}_{\b(1234)}=
{1\over\sqrt{\D (1234)}}\sum_M C^{N(1234) M}_{\b(1234)} \,.
\ee
$M$ denotes the
simple irreps whose singlets appear in the tensor product of four simple 
irreps, and $\D$ is the dimension of that subspace. By inserting (\ref{ghfe}) 
into $S_4$ one obtains
\be
S_{k} = \half \sum_{N,\a} \left| \f^{N(1234)}_{\a(1234)}\right|^2 
\quad,
\ee
while for the interaction part we get
\be
S_{v}= {\l\over 5!}\sum_{N,\a }
\f^{N(1234)}_{\a(1234)}\,\f^{N(4567)}_{\a(4567)}\,
\f^{N(7389)}_{\a(7389)}
\,\f^{N(96210)}_{\a(96210)}\f^{N(10851)}_{\a(10851)}\,
\tilde\cv (N_1 \cdots N_{10})\,,\label{bcia}
\ee
where $\tilde\cv$ is the normalized BC vertex
\be
\tilde\cv = {\cv (N_1 \cdots N_{10})\over \sqrt{\D (1234)\cdots \D (10851)}}
\quad.
\ee 
 
Hence the FD of the field theory defined by (\ref{bcia}) will be given by 
the closed and open 5-valent graphs, as in the topological case, but the
propagator and the vertex factors will be different. A vacuum FD $\G$ will 
reproduce the state sum (\ref{bcss}) for a triangulation $\ct$ whose dual 
one-complex is $\G$. However, now one expects that the regularized sum
will depend on the triangulation $\ct$, 
since $Z(\ct)$ does not have the properties of a topological invariant. 

Note that there exists a modification of the BC field theory
action which gives finite FD in the Euclidean case without using the
quantum group as a regulator \cite{pr,per}. The modified action has the same
kinetic term as the topological action ($G$ invariance only), while the 
interaction term is the same as in the BC case ($G$ and $H$ invariance). 
When expanded in the Fourier modes, one obtains 
\be
S_k = \half \sum_{N,M,\a} (d_{N_1} ... d_{N_4} )^2 
\left| \f^{N(1234)M}_{\a(1234)}\right|^2 
\label{prbca}
\ee
for the kinetic term, while 
\be
S_v = {\l\over 5!}\sum_{N,M,\a }
\f^{N(1234)M_1}_{\a(1234)}\,\f^{N(4567)M_2}_{\a(4567)}\,
\f^{N(7389)M_3}_{\a(7389)}
\,\f^{N(96210)M_4}_{\a(96210)}\f^{N(10851)M_5}_{\a(10851)}\,
\tilde\cv (N_1 \cdots N_{10})\,, \nonumber
\ee
is the interaction term.
The appearance of the factor $(d_{N_1} ... d_{N_4} )^2 $ 
in the kinetic term
gives a damping factor $(d_{N_1} ... d_{N_4} )^{-2}$ 
in the propagator which insures the finiteness.

As far as the  FD with external legs are concerned, they will correspond
to triangulations of a 4-manifold with boundaries, 
where each boundary contains more then one 4-simplex. The boundary complexes
are then interpreted as 3d complexes, corresponding to triangulations
of boundary 3-manifolds $\S_i$. These FD will be relevant for the
transition amplitude from a state on a 3-manifold $\S_1$ to a state on a
3-manifold $\S_2$, where
$\pr M = \S_1 \cup \S_2$. However, we need to construct first the spin net 
state associated to
a triangulation of a 3-manifold $\S$ representing a boundary of $M$. 
This will be given by a straightforward generalization of the results from 
the lower dimensions. 

Consider
a 4-valent connected closed graph $\g$ which corresponds to a 
triangulation of $\S$, i.e. $\g$ is a dual one complex to $\ct(\S)$, 
where now $\ct(\S)$ is considered as a 3d simplical complex. 
For example, a 4-simplex is dual to a pentagram. A more complicated example 
is a 3d simplical complex consisting of
8 tetrahedrons. The dual graph will have eight 4-valent 
vertices and 16 edges.

Given a triangulation $\ct(\S)$ we can construct a functional
$\F$ by integrating a product of $\vf_i (g_j)$ over $G^n$, 
where $n$ is the number of triangles in $\ct (\S)$ and $\vf_i (g_j)$
is associated to the tetrahedron in $\ct (\S)$ whose triangles are colored 
with the group elements $g_j$, i.e.
\be
\F = \int_{G^n} d^n g \, \prod_{t\in \ct(\S)} 
\vf ( g_{f_1(t)},g_{f_2 (t)},g_{f_3 (t)},g_{f_4 (t)}) \quad.
\ee
By Fourier expanding it, we get
\be
\F = \sum_{N,\a}\f^{N(1234)}_{\a(1234)}
\cdots \f^{N(n_1 n_2 n_3 n_4)}_{\a(n_1 n_2 n_3 n_4)}
\cv_\g (N_1 N_2  \cdots  N_{n_4} )\,,
\ee
where the vertex $\cv_\g (N_1 \cdots  N_{n_4})$ is given by (\ref{snv})
and $\g$ is the four-valent graph associated to $\ct (\S)$, i.e. a dual
one-complex to $\ct(\S)$. Hence we take the operator 
\bea
\F (\g,N ) &=& {1\over N_0 (N)}\sum_\a \f^{N(1234)}_{\a(1234)}
\cdots \f^{N(n_1 n_2 n_3 n_4)}_{\a(n_1 n_2 n_3 n_4)} \nonumber\\
&=&{1\over N_0 (N)}\sum_\a \prod_{v\in\g}
\f^{N(e_1(v)... e_4(v))}_{\a(e_1(v)... e_4(v))} \label{ftsn}
\eea
to represent the spin net given by the graph $\g$ whose edges are labeled 
with the simple 
irreps $N_1,...,N_{n_4}$. In the case of the Perez-Rovelli model, one obtains
\be
\F (\g,N ) = {1\over N_0 (N)}\sum_{M,\a} \prod_{v\in\g}
\f^{N(e_1(v)...e_4(v))M_v}_{\a(e_1(v)...e_4(v))}\,. \label{prsn}
\ee

The amplitude
\be
A_{12}(n) = \langle \F (\g_2 ,N_2)(S_v)^n \F (\g_1 ,N_1 )
\rangle 
\quad,\label{grta}
\ee
will be interpreted as the amplitude for a transition from the state 
$\ket{\F (\g_1 ,N_1 )}$ to the state $\ket{\F (\g_2 ,N_2 )}$
in $n$ units of the covariant time $T$. This will be a finite sum of FD, 
consisting of $(v_1 , v_2)$ FD 
with $n$ vertices and $v_1 + v_2$ external legs, where 
$v_i$ is the number of the vertices in the spin net $\g_i$. Each FD would give
a 4d analog of the single-complex Ponzano-Regge amplitude, if we choose a 
normalization $N_0 (N)=\sqrt{\prod_k d(N_k)}$.

Note that the amplitude (\ref{grta}) will be finite in the Euclidean
Perez-Rovelli model. Therefore this would be en example of a finite
4d quantum gravity theory with local degrees of freedom, provided that one uses
the covariant time interpretation. In the Lorentzian
case it is not known whether the Perez-Rovelli model is finite, but it is 
plausible that it may be finite. Alternatively, one can try the quantum 
Lorentz group regularization. 

As in the 3d case, the field theory spin nets $\F$ can be considered as 
generalizations of the canonical quantization spin nets $\Psi$ \cite{csn} 
to situations
when the topology of the spacetime manifold is not fixed. Also, due to 
the non-topological nature of 4d gravity, 
the labels are not the same anymore. Covariant spin nets have simple
$SO(4)$ or $SL(2,C)$ irreps as labels, while the canonical spin nets have
the $SU(2)$ labels. In order to make a closer link between these, one would
have to discretize the canonical spin nets, as in the 3d case, but in the
4d case one would have to solve the diffeomorphism and the Hamiltonian
constraints. Furthermore, one would have to find a time variable in order to 
compute the 
transition amplitudes. A great advantage of the covariant approach, which
is related to the fact that its a path-integral quantization, is that
one does not have to do all these non-trivial steps, and one can calculate  
directly the transition amplitudes. 

\section{Conclusions}

We have demonstrated that there is a significant benefit if one completes the
Feynman diagram picture of the transition amplitudes for the state-sum models
of quantum gravity by introducing the Fourier mode operators and the
corresponding Fock space, so that the transition amplitudes can be understood 
as matrix elements of an evolution operator, in exact analogy with the particle
field theory case. The spin network operators are constructed from 
the products of the field Fourier modes, such that each mode 
corresponds
to a vertex of the spin net graph, and then one averages over the spin 
components in a group-theory invariant way. The corresponding state
can be thought of as a state of a universe whose space is discrete. This 3d 
space is given by the dual 3-complex to the spin net graph $\g$, and the 
quantum states are determined by the colorings of the triangles  
with the irreps $\L$. 

We have considered only the transition amplitudes for the single-universe 
states. Since one can act on the Fock space
vacuum with more than one spin net operator, one can obtain the 
multi-universe states, and the corresponding transition amplitudes.
Hence our quantum field theory of spin networks is also a quantum field theory
of discrete universes. This QFT is then a discrete realization of the third 
quantization of gravity.   

Note that the expressions for the amplitudes we have written are mostly
formal, and they need a regularization for the infinite sums over the
irreps of $G$. The standard approach is to use the quantum group formalism,
which is a group invariant way of putting a cut off on $\L$. Alternatively, 
the Euclidean BC model can be regularized via Perez-Rovelli reformulation. 
Hence in this
model the amplitudes for finite time intervals will be finite. The same could
apply for the Lorentzian version of the Perez-Roveli model \cite{prl}, 
provided one
proves the finiteness of an arbitrary Feynman diagram. If this turns out to be 
true, one would have a well defined quantum theory of 4d Lorentzian gravity.
Then the main task would be to examine the semi-classical states of this theory.
Given the discrete nature of the theory, one expects that such states would 
be described by spin nets of large number of vertices, since large 
triangulations represent good approximations of smooth manifolds.

The key ingredient in making the quantum field theory of spin nets well 
defined is the introduction of the discrete time variable $T$ and the 
corresponding evolution operator $U(T)$. Since $U(T)= (S_v)^T$ one can 
define the transition amplitudes for finite time intervals, which are given
by the sum of FD with $T$ vertices, or equivalently, by 
the sum of partition functions for triangulations with $T$ simplices. 
In this way one avoids
the infinite sum of the single-complex amplitudes, or Feynman diagrams, 
for all possible complexes interpolating between 
the boundary complexes, which is often invoked in the literature as the
physical amplitude. We have demonstrated that this is not necessary, since
this prescription is related to the evolution operator $U = \exp (iS_v)$,
and this operator is not natural in the covariant quantization formulation
of theories without a fixed spacetime background.

The construction of the covariant time evolution operator   
mirrors the physical intuition that summing over all possible 
triangulations of the spacetime manifold corresponds to 
including metrics 
which give arbitrary large geodesic distance between the initial and the final 
boundary. Therefore one should divide this sum into parts corresponding to
fixed time-like geodesic lengths. These partial sums still contain infinitely
many triangulations. Hence a better time variable, i.e. a variable which 
gives finite 
partial sums, is the discretized analog of the spacetime volume, 
which is the number of the $D$-simplices in a simplical complex representing
the spacetime. Also, this interpretation is natural for the compact universes.

In order to gain a better understanding of the covariant time, one would have
to see what is its relation to canonical time variables\footnote{These are 
time variables in the canonical formulation of GR \cite{ct}. In this context 
a typical canonical time variable would be the volume of a spatial section, 
which would correspond to the number of vertices in the spin net.}, as well as
a relation to matter clock variables. This analysis will be also important for
understanding the meaning of the non-unitarity of the covariant time evolution
operator.

If one takes the covariant evolution operator as the fundamental object,
this would mean abandoning the law of unitary evolution of states in quantum
mechanics (QM). However, this may not be a bad thing in this context, since 
we are dealing with the wavefunction of 
the Universe, which as a concept is not well defined in the standard QM due to
the fact that there is no outside observer. Also, it could provide a 
mechanism for the the matter wavefunction collapse, which would be a concrete
realization of gravity driven 
wavefunction collapse \cite{pen}. Also note that the QFT formulation of 
spin nets
is also a good arena for exploring other interpretation schemes of
QM for quantum gravity \cite{ct}.

In order to explore all these ideas, as well as in order to have a complete 
theory, one would need a simplical field theory formulation of matter coupled 
to gravity.

\section*{Acknowledgements} 

I would like to thank L. Crane, R. Picken and J. Mourao for stimulating
discussions. Work supported by the grant 
PRAXIS/BCC/18981/98 from the Portugese Foundation for Science and Technology
(FCT).

\end{document}